\documentclass[sigconf, manuscript]{acmart}
\AtBeginDocument{%
  }

\setcopyright{none}
\renewcommand\footnotetextcopyrightpermission[1]{}
\acmConference{}{}{}
\acmISBN{}

\copyrightyear{2024}
\acmYear{2024}
\acmDOI{XXXXXXX.XXXXXXX}

\usepackage{tabularx}
\usepackage{multirow}
\usepackage{diagbox}

\begin{document}

\title{Exploring Capabilities of Time Series Foundation Models in Building Analytics}

\author{Xiachong Lin}
\orcid{0009-0009-6762-8365}
\affiliation{%
  \institution{University of New South Wales}
  \city{Sydney}
  \state{NSW}
  \country{Australia}
}
\email{dawn.lin@student.unsw.edu.au}

\author{Arian Prabowo}
\affiliation{%
  \institution{University of New South Wales}
  \city{Sydney}
  \state{NSW}
  \country{Australia}}
\email{arian.prabowo@unsw.edu.au}

\author{Imran Razzak}
\affiliation{%
  \institution{University of New South Wales}
  \city{Sydney}
  \state{NSW}
  \country{Australia}}
\email{imran.razzak@unsw.edu.au}

\author{Hao Xue}
\affiliation{%
  \institution{University of New South Wales}
  \city{Sydney}
  \state{NSW}
  \country{Australia}}
\email{hao.xue1@unsw.edu.au}

\author{Matthew Amos}
\affiliation{%
  \institution{CSIRO Energy Centre}
  \city{Newcastle}
  \state{NSW}
  \country{Australia}}
\email{matt.amos@csiro.au}

\author{Sam Behrens}
\affiliation{%
  \institution{CSIRO Energy Centre}
  \city{Newcastle}
  \state{NSW}
  \country{Australia}}
\email{sam.behrens@csiro.au}

\author{Flora D. Salim}
\affiliation{%
  \institution{University of New South Wales}
  \city{Sydney}
  \state{NSW}
  \country{Australia}}
\email{flora.salim@unsw.edu.au}

\renewcommand{\shortauthors}{}

\begin{abstract}

The growing integration of digitized infrastructure with Internet of Things (IoT) networks has transformed the management and optimization of building energy consumption. By leveraging IoT-based monitoring systems, stakeholders such as building managers, energy suppliers, and policymakers can make data-driven decisions to improve energy efficiency. However, accurate energy forecasting and analytics face persistent challenges, primarily due to the inherent physical constraints of buildings and the diverse, heterogeneous nature of IoT-generated data. In this study, we conduct a comprehensive benchmarking of two publicly available IoT datasets, evaluating the performance of time series foundation models in the context of building energy analytics. Our analysis shows that single-modal models demonstrate significant promise in overcoming the complexities of data variability and physical limitations in buildings,  with future work focusing on optimizing multi-modal models for sustainable energy management. 

\end{abstract}


\begin{teaserfigure}
  \includegraphics[width=\textwidth]{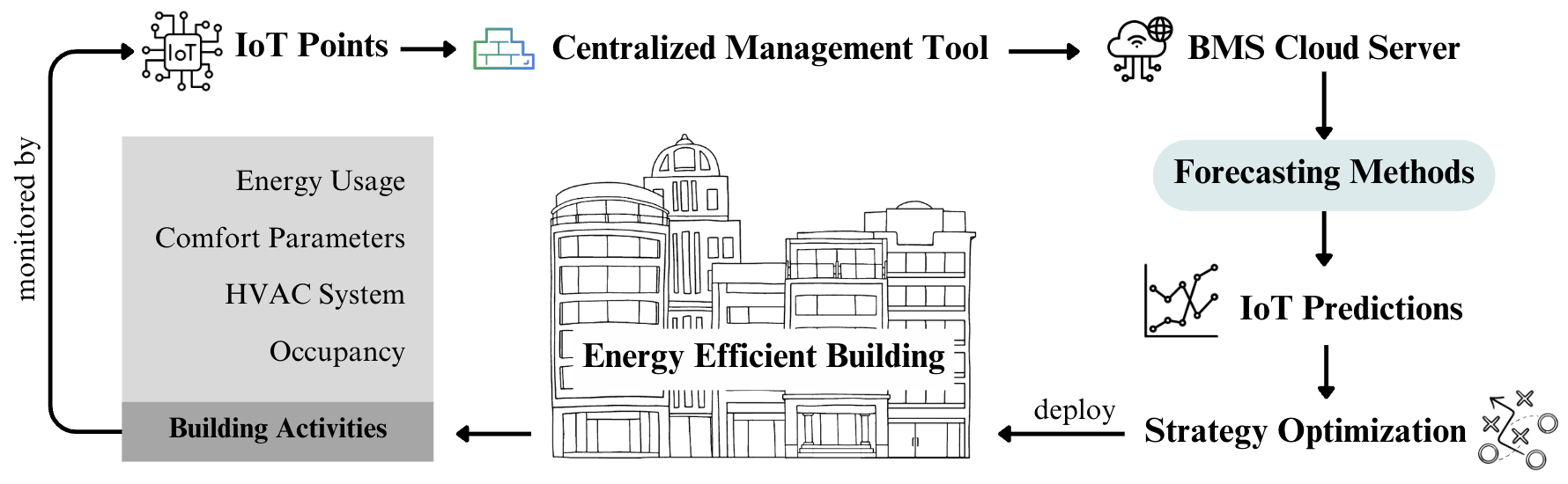}
  \caption{A sample application for building energy control utilizing IoT-generated data. Forecasting models access IoT data streams from a cloud-based Building Management System(BMS) to generate predictions. These predictive results form the basis for energy management and optimization, aiding in achieving load-shifting objectives.}
  \Description{Enjoying the baseball game from the third-base
  seats. Ichiro Suzuki preparing to bat.}
  \label{fig:teaser}
\end{teaserfigure}


\maketitle

\section{Introduction}
The deployment of AI in digital infrastructure presents significant opportunities for enhancing energy efficiency and advancing net-zero strategies. Traditional building energy research relies on post-processed datasets, which typically include aggregated load data and weather station-monitored temperature—referred to as rough granularity data. While these higher-level data offer convenience for data scientists by providing a broad overview, they compromise the granularity necessary for a detailed description of activities. The introduction of the BLDG59\cite{luo2022three} and BTS datasets\cite{prabowo2024bts} based on the centralized management tool, Brick Schema\cite{balaji2018brick}, addresses this limitation by offering comprehensive IoT sensing and metering point data, enabling the exploration of detailed usage patterns in digitalized buildings. However, the increased data texture complicates the building activity modeling due to the varied architecture preferences, energy efficiency measures, local climate conditions, usage patterns, and engineering practices, all contributing to heterogeneous sensing data among buildings\cite{lin2024gap}. Accurate cross-building forecasting using IoT point data necessitates models that can adapt to unseen physical constraints and variations in ontology usage.



Foundation models, which utilize large, pre-trained models as the base for fine-tuning specific tasks, are receiving increasing attention in time series analysis due to their capabilities in generalization. Existing time series foundation models are typically summarized into two categories. Prompt-based methods, such as PromptCast\cite{xue2023promptcast} and LLMTime \cite{gruver2024large}, typically convert time series to sentences using a manually defined template and feed to language models. Methods following this logic benefit from the easy implementation, but specific prompt templates are required based on targeted tasks and datasets. 
Tuning-based models typically employ fine-tuning techniques for LLMs to adapt the time series inputs. One-Fits-All (GPT4TS)\cite{zhou2023one} and LLM4TS\cite{chang2023llm4ts} approach multivariate time series by treating them as univariate sequences, which are then segmented into patches. These patches are encoded using specific encoders, concatenated, and subsequently fed into the pre-trained LLMs. To enhance forecasting by incorporating domain knowledge, GPT4MTS\cite{jia2024gpt4mts} introduces a prompting template that combines time series data with textual prompts. TimeLLM\cite{jin2023time} initially proposes the cross-modal attention mechanism, which aligns the time series with pre-train LLM's word embedding. Inspired by this work, LLaTA\cite{liu2024taming} designed a dual-branch architecture to process the textual and temporal modalities. Each branch fine-tunes a GPT-2 backbones with the Low-rank adaptation technique (LoRA)\cite{hu2021lora} to adapt to the task, aligning the distillation knowledge and temporal embedding by reducing the distribution discrepancy between the modalities. 

TS foundation models are typically evaluated on well-processed datasets, which differ significantly from real-world building IoT datasets. Building IoT contains branches of information about energy usage, comfort parameters, and environmental factors, but at the same time, the complex dependencies and data heterogeneity among the IoT network present challenges for machine learning and deep learning techniques. In the context of energy-efficient digital infrastructures, the application of TS models still needs extensive experimentation to validate their effectiveness. This comparative study aims to assess the performance of TS foundation models in the building energy domain, benchmarking them against traditional machine learning and deep learning approaches. Additionally, the study explores architectural ablations of TS foundation models to understand their strengths and limitations. The remainder of the paper is structured as follows: Section \ref{sec:method} evaluation methodology adopted by this benchmark study. Section \ref{sec:results} presents the results, followed by a discussion of the findings and future research directions in Section \ref{sec:discussion}. Section \ref{sec:conclusion} provides a concise conclusion.



\section{Methodology}
\label{sec:method}


\subsection{Datasets}
This study employs the following two public datasets:

\subsubsection{BTS-B\cite{prabowo2024bts}} The BTS-B contains 730 IoT points data, including 264 sensors. $nUnique/nSample\geq\sigma$ is used to filter the sensor data streams,
105 streams are chosen for further processing. Each time series is resampled to 10-min, where missing values are filled with second-order polynomial interpolation.
Records from 00:00:00 01/06/2022 to 00:00:00 01/08/2022 are used for training, whereas records from 00:00:00 01/08/2022 to 00:00:00 01/09/2022 are used for testing. 

\subsubsection{BLDG59\cite{luo2022three}} Following the similar processing procedure as BTS-B, records for 49 selected sensors in BLDG59 from 00:00:00 01/06/2019 to 00:00:00 01/08/2019 are used for training. Records from 00:00:00 01/08/2019 to 00:00:00 01/09/2019 are used for testing. The original data is downsampled by mean values and interpolated upsampled by second-order polynomials for multi-timestep alignment, such that the granularity is set to 10-min. In both of the cases, $\sigma=0.1$.

\subsection{Experimental Settings}
The train/validation ratio is set to 7/3. 
Adam optimization with a 1e-3 learning rate guides the learning process. 
Training epochs are set to 100 while early stopping with 10-epoch patience is employed. All the models are respectively fed with 1-day historical observations and forecasts on multiple time steps, where $S=144$ and $H\in\{12, 48, 96, 144, 432, 1008\}$, indicating 2-hour, 8-hour, 16-hour, 1-day, 3-day, and 1-week ahead forecasting. The average scores across all the forecasting horizons are computed. 
The weights are updated using the Adam optimizer with a model-specific learning rate. EarlyStop with a patience equal to 10 is used for the experiments. The batch size of each model is adjusted to fit the GPU utilization rate. All experiments are completed on V100 or H100 GPUs.

\subsection{Baselines\&Evaluation} The following models are employed by this comparative study: 
\textit{\textbf{Machine Learning Methods}} RandomForest\cite{breiman2001random} and XGBoost\cite{chen2016xgboost}. \textit{\textbf{Deep Learning Models}} DLinear\cite{zeng2023transformers}, PatchTST\cite{nie2022time}, Informer\cite{zhou2021informer} and iTransformer\cite{liu2023itransformer}. \textit{\textbf{Time Series Foundation Models}} LLaTA\cite{liu2024taming}, One-Fits-All\cite{zhou2023one} and TimeLLM\cite{jin2023time}, all TS foundation models share the same backbone GPT-2 model. Additionally, due to memory constraints, we applied offline PCA with n\_component=500, as\cite{liu2024taming}, to reduce the dimensions of the word embeddings for TimeLLM, rather than using the original embeddings outlined in the original paper.
All baseline experiments are conducted using the zero-shot settings, where the model trained on one dataset with 2-month records is directly applied to forecast the following 1-month records on another dataset without any retraining. 
\textit{\textbf{Metrics}}
Mean absolute error (MAE), mean square error (MSE), and symmetric mean absolute percentage error (SMAPE) are employed to evaluate the model performance. 
\section{Results}
\label{sec:results}
\begin{table*}[htb]
    \centering
    \small
    \begin{tabular}{l|l|ccc|ccc}
    \toprule
    \multirow{2}{*}{\diagbox{Train}{Test}} & \multirow{2}{*}{Method} & \multicolumn{3}{c}{BTS-B} & \multicolumn{3}{|c}{BLDG59} \\ \cmidrule{3-8}
        & & MAE & MSE & SMAPE & MAE & MSE & SMAPE \\
        \hline
    \multirow{8}{*}{BTS-B} 
    & RandomForest & \textcolor{red}{$0.542\pm0.088$} & $0.668\pm0.126$ & $50.034\pm8.671$ & $0.575\pm0.130$ & $0.658\pm0.210$ & $50.734\pm10.523$ \\
    & XGBoost & $0.600\pm0.054$ & \textcolor{red}{$0.659\pm0.095$} & $55.289\pm5.823$ & $0.618\pm0.089$ & $0.674\pm0.170$ & $55.914\pm7.989$ \\
    & DLinear & $0.552\pm0.079$ & $0.669\pm0.111$ & $50.039\pm7.863$ & $0.560\pm0.132$ & \textcolor{red}{$0.630\pm0.222$} & $48.686\pm9.938$ \\
    & PatchTST & $0.551\pm0.090$ & $0.704\pm0.152$ & $47.463\pm6.370$ & $0.589\pm0.156$ & $0.749\pm0.318$ & $47.383\pm8.841$ \\
    & Informer & $0.596\pm0.107$ & $0.752\pm0.163$ & $53.670\pm11.775$ & $0.626\pm0.135$ & $0.741\pm0.215$ & $54.364\pm12.574$ \\
    & iTransformer & $0.567\pm0.068$ & $0.730\pm0.132$ & $47.577\pm4.358$ & $0.653\pm0.133$ & $0.830\pm0.282$ & $52.793\pm7.865$ \\
    & LLaTA & $0.809\pm0.047$ & $1.187\pm0.099$ & $61.976\pm2.636$ & $0.834\pm0.090$ & $1.224\pm0.235$ & $61.473\pm3.797$  \\
    & One-Fits-All & $0.552\pm0.089$ & $0.696\pm0.151$ & \textcolor{red}{$44.873\pm6.094$} &  \textcolor{red}{$0.558\pm0.168$} & $0.733\pm0.355$ & \textcolor{red}{$43.530\pm8.991$} \\
    & TimeLLM & $0.559\pm0.092$ & $0.695\pm0.149$ & $46.569\pm6.767$ & $0.571\pm0.176$ & $0.734\pm0.364$ & $45.589\pm9.971$  \\
    \hline 
    \multirow{8}{*}{BLDG59}
    & RandomForest & $0.594\pm0.093$ & $0.710\pm0.113$ & $53.538\pm10.753$ & $0.554\pm0.133$ & $0.619\pm0.204$ & $49.806\pm11.700$ \\
    & XGBoost & $0.627\pm0.066$ & $0.702\pm0.107$ & $59.375\pm8.307$ & $0.606\pm0.095$ & $0.645\pm0.174$ & $56.251\pm9.370$ \\
    & DLinear & $0.565\pm0.091$ & \textcolor{red}{$0.687\pm0.114$} & $51.454\pm10.240$ & \textcolor{red}{$0.547\pm0.132$} & \textcolor{red}{$0.603\pm0.201$} & $48.572\pm11.267$\\
    & PatchTST & $0.581\pm0.103$ & $0.760\pm0.163$ & $49.852\pm7.749$ & $0.566\pm0.169$ & $0.711\pm0.332$ & $45.872\pm10.088$ \\
    & Informer & $0.627\pm0.111$ & $0.811\pm0.185$ & $56.527\pm11.029$ & $0.604\pm0.159$ & $0.721\pm0.281$ & $52.833\pm12.943$ \\
    & iTransformer & $0.654\pm0.124$ & $0.984\pm0.269$ & $50.842\pm7.450$ & $0.582\pm0.177$ & $0.760\pm0.361$ & $45.753\pm10.516$ \\
    & LLaTA & $0.809\pm0.047$ & $1.188\pm0.100$ & $62.032\pm2.688$ & $0.834\pm0.091$ & $1.226\pm0.237$ & $61.443\pm3.772$ \\
    & One-Fits-All & \textcolor{red}{$0.563\pm0.090$} & $0.706\pm0.138$ & \textcolor{red}{$45.824\pm6.855$} & $0.551\pm0.172$ & $0.710\pm0.351$ & \textcolor{red}{$43.165\pm9.643$} \\
    & TimeLLM & $0.570\pm0.092$ & $0.707\pm0.144$ & $47.960\pm7.389$ & $0.564\pm0.180$ & $0.718\pm0.368$ & $45.259\pm10.468$ \\
    \bottomrule
    \end{tabular}
    \caption{Average results (mean$\pm$std) on $\{$12, 48, 96, 144, 432, 1008$\}$ time-step ahead forecasting. The input sequence lengths of all the baselines are set to 144. The best performance on each column is highlighted in \textcolor{red}{red}.}
    \label{tab:main_result}
\end{table*}
\subsection{Results of Building IoT Forecasting}
Table\ref{tab:main_result} indicates the result of the comparative study. Each model trained on a specific dataset is evaluated on the same and unseen datasets to assess its generalization performance across different data distributions. One-Fits-All achieves the lowest SMAPE across all train/test configurations, consistently delivering superior performance in percentage-based error, and secures the top position 6 times across all metrics, highlighting its robustness in both relative and absolute error handling. LLaTA demonstrates consistently weaker performance across most metrics, particularly higher errors in both MSE and SMAPE across the two datasets. While TimeLLM performs reasonably well in terms of cross-building forecasting, One-Fits-All outperforms it with a notable margin of 3.642\%, 4.520\%, 4.454\%, and 4.627\% in the BTS-B/BTS-B, BTS-B/BLDG59, BLDG59/BTS-B, and BLDG59/BLDG59 train/test configurations, respectively.
\subsection{Ablation Studies for TS Foundation Models}
To assess the effectiveness of LLMs in foundation models for building IoT, we adopt an ablation framework designed by Tan et al.\cite{tan2024language}. This framework introduces modifications to the original models by either removing the LLM (w/o LLM), replacing it with an attention layer (LLM2Attn), or substituting it with a Transformer layer (LLM2Trsf). The ablation results are presented in Tables \ref{tab:llata}-\ref{tab:timellm}. Tan et al. argue that LLMs may not offer significant advantages for time series forecasting, and our experiments on LLaTA and TimeLLM generally support this view. However, a notable observation is that One-Fits-All outperforms its ablations in 4 out of 6 metrics for cross-building scenarios. This suggests that One-Fits-All, as a single-modal foundation model, is capable of addressing the challenges posed by building heterogeneity. Multi-modal and multi-source data employed by LLaTA and TimeLLM may not integrate seamlessly, potentially confusing the model and leading to suboptimal performance. In contrast, foundation models focus on single-modal inputs, avoiding the complexity and potential overhead of aligning multiple data types, which can optimize their architecture for that specific task. 

\begin{table*}[htb]
    \centering
    \small
    \begin{tabular}{l|l|ccc|ccc}
    \toprule
    \multirow{2}{*}{\diagbox{Train}{Test}} & \multirow{2}{*}{Method} & \multicolumn{3}{c}{BLDG59} & \multicolumn{3}{|c}{BTS-B} \\ \cmidrule{3-8}
        & & MAE & MSE & SMAPE & MAE & MSE & SMAPE \\
        \hline
    \multirow{5}{*}{BLDG59}
    & LLM2Attn & \textcolor{red}{$0.816\pm0.091$} & \textcolor{red}{$1.168\pm0.234$} & $61.484\pm3.836$ & $0.788\pm0.046$ & \textcolor{red}{$1.122\pm0.096$} & $62.235\pm2.718$ \\
    & LLM2Trsf & $0.818\pm0.088$ & $1.173\pm0.229$ & $61.578\pm3.752$ & $0.790\pm0.044$ & $1.127\pm0.090$ & $62.331\pm2.644$ \\
    & w/o LLM2 & $0.816\pm0.090$ & $1.169\pm0.232$ & $61.491\pm3.824$ & \textcolor{red}{$0.788\pm0.045$} & $1.123\pm0.094$ & $62.238\pm2.712$  \\
    & LLaTA & $0.834\pm0.091$ & $1.226\pm0.237$ & \textcolor{red}{$61.443\pm3.772$} & $0.809\pm0.047$ & $1.188\pm0.100$ & \textcolor{red}{$62.032\pm2.688$}  \\
    \hline 
    \multirow{5}{*}{BTS-B}
    & LLM2Attn & \textcolor{red}{$0.815\pm0.090$} & \textcolor{red}{$1.165\pm0.232$} & $61.516\pm3.862$ & \textcolor{red}{$0.787\pm0.045$} & \textcolor{red}{$1.121\pm0.095$} & $62.178\pm2.666$ \\
    & LLM2Trsf & $0.817\pm0.088$ & $1.170\pm0.227$ & $61.610\pm3.777$ & $0.789\pm0.043$ & $1.127\pm0.089$ & $62.275\pm2.592$ \\
    & w/o LLM2 & $0.816\pm0.089$ & $1.167\pm0.230$ & $61.523\pm3.851$ & $0.787\pm0.044$ & $1.123\pm0.093$ & $62.181\pm2.660$ \\
    & LLaTA & $0.834\pm0.090$ & $1.224\pm0.235$  & \textcolor{red}{$61.473\pm3.797$} & $0.809\pm0.047$ & $1.187\pm0.099$ & \textcolor{red}{$61.976\pm2.636$} \\
    \bottomrule
    \end{tabular}
    \caption{Average results (mean$\pm$std) on $\{$12, 48, 96, 144, 432, 1008$\}$ time-step ahead forecasting on LLaTA. The input sequence lengths of all the baselines are set to 144. The best performance on each block is highlighted in \textcolor{red}{red}.}
    \label{tab:llata}
\end{table*}
\begin{table*}[htb]
    \centering
    \small
    \begin{tabular}{l|l|ccc|ccc}
    \toprule
    \multirow{2}{*}{\diagbox{Train}{Test}} & \multirow{2}{*}{Method} & \multicolumn{3}{c}{BLDG59} & \multicolumn{3}{|c}{BTS-B} \\ \cmidrule{3-8}
        & & MAE & MSE & SMAPE & MAE & MSE & SMAPE \\
        \hline
    \multirow{5}{*}{BLDG59}
    & LLM2Attn & $0.553\pm0.176$ & \textcolor{red}{$0.713\pm0.357$} & $43.626\pm10.001$ & $0.579\pm0.102$ & $0.725\pm0.156$ & $47.605\pm7.815$\\
    & LLM2Trsf & \textcolor{red}{$0.550\pm0.180$} & $0.716\pm0.366$ & \textcolor{red}{$43.074\pm10.177$} & $0.582\pm0.102$ & $0.737\pm0.157$ & $47.481\pm7.923$ \\
    & w/o LLM2 & $0.555\pm0.171$ & $0.725\pm0.349$ & $43.450\pm9.553$ & $0.570\pm0.087$ & $0.727\pm0.130$ & $45.982\pm6.773$ \\
    & One-Fits-All & $0.551\pm0.172$ & 0.717$\pm0.351$ & $43.165\pm9.643$ & \textcolor{red}{$0.563\pm0.090$} & \textcolor{red}{$0.706\pm0.138$} & \textcolor{red}{$45.824\pm6.855$} \\
    \hline 
    \multirow{5}{*}{BTS-B}
    & LLM2Attn & $0.565\pm0.166$ & \textcolor{red}{$0.730\pm0.346$} & $44.659\pm9.058$ & $0.552\pm0.093$ & $0.692\pm0.159$ & $45.498\pm6.237$ \\
    & LLM2Trsf & $0.569\pm0.170$ & $0.739\pm0.348$ & $45.061\pm9.480$ & \textcolor{red}{$0.547\pm0.094$} & \textcolor{red}{$0.685\pm0.156$} & $45.372\pm6.546$ \\
    & w/o LLM2 & \textcolor{red}{$0.557\pm0.170$} & $0.730\pm0.355$ & $43.538\pm9.082$ & $0.563\pm0.088$ & $0.714\pm0.141$ & $45.505\pm6.273$ \\
    & One-Fits-All & $0.558\pm0.168$ & $0.733\pm0.355$ & \textcolor{red}{$34.755\pm7.363$} & $0.552\pm0.089$ & $0.696\pm0.151$ & \textcolor{red}{$38.933\pm5.940$} \\
    \bottomrule
    \end{tabular}
    \caption{Average results (mean$\pm$std) on $\{$12, 48, 96, 144, 432, 1008$\}$ time-step ahead forecasting on One-Fits-All(GPT4TS). The input sequence lengths of all the baselines are set to 144. The best performance on each block is highlighted in \textcolor{red}{red}.}
    \label{tab:ofa}
\end{table*}

\begin{table*}[htb]
    \centering
    \small
    \begin{tabular}{l|l|ccc|ccc}
    \toprule
    \multirow{2}{*}{\diagbox{Train}{Test}} & \multirow{2}{*}{Method} & \multicolumn{3}{c}{BLDG59} & \multicolumn{3}{|c}{BTS-B} \\ \cmidrule{3-8}
        & & MAE & MSE & SMAPE & MAE & MSE & SMAPE \\
        \hline
    \multirow{5}{*}{BLDG59}
    & LLM2Attn & $0.552\pm0.188$ & $0.709\pm0.368$ & $44.465\pm11.690$ & \textcolor{red}{$0.562\pm0.100$} & $0.698\pm0.153$ & $47.880\pm8.702$ \\
    & LLM2Trsf & \textcolor{red}{$0.547\pm0.188$} & \textcolor{red}{$0.707\pm0.369$} & \textcolor{red}{$43.919\pm11.657$} & $0.563\pm0.103$ & $0.698\pm0.150$ & $47.793\pm9.178$ \\
    & w/o LLM2 & $0.564\pm0.181$ & $0.724\pm0.367$ & $45.002\pm10.877$ & $0.564\pm0.090$ & \textcolor{red}{$0.698\pm0.135$} & \textcolor{red}{$47.190\pm7.829$} \\
    & TimeLLM & $0.564\pm0.180$ & $0.718\pm0.368$ & $45.259\pm10.468$ & $0.570\pm0.092$ & $0.707\pm0.144$ & $47.960\pm7.389$ \\
    \hline
    \multirow{5}{*}{BTS-B}
    & LLM2Attn & \textcolor{red}{$0.562\pm0.183$} & $0.736\pm0.368$ & \textcolor{red}{$44.696\pm10.990$} & $0.543\pm0.098$ & $0.679\pm0.158$ & $45.778\pm7.688$ \\
    & LLM2Trsf & $0.568\pm0.184$ & $0.746\pm0.364$ & $45.083\pm11.251$ & \textcolor{red}{$0.538\pm0.097$} & \textcolor{red}{$0.671\pm0.154$} & \textcolor{red}{$45.348\pm7.757$} \\
    & w/o LLM2 & $0.567\pm0.179$ & \textcolor{red}{$0.730\pm0.365$} & $45.480\pm10.601$ & $0.559\pm0.092$ & $0.691\pm0.143$ & $46.354\pm7.646$ \\
    & TimeLLM & $0.571\pm0.176$ & $0.734\pm0.364$ & $45.589\pm9.971$ & $0.559\pm0.092$ & $0.695\pm0.149$ & $46.569\pm6.767$ \\
    \bottomrule
    \end{tabular}
    \caption{Average results (mean$\pm$std) on $\{$12, 48, 96, 144, 432, 1008$\}$ time-step ahead forecasting on TimeLLM. The input sequence lengths of all the baselines are set to 144. The best performance on each block is highlighted in \textcolor{red}{red}.}
    \label{tab:timellm}
\end{table*}

\section{Discussion}
\label{sec:discussion}

The primary challenge in modeling building activity using IoT-generated data lies in addressing the heterogeneity of buildings, each of which has unique energy consumption patterns influenced by factors such as structural design, insulation quality, HVAC systems, and occupancy behavior. Furthermore, the diversity in energy systems and load management practices adds complexity, making accurate forecasting in heterogeneous building environments more difficult. The ideal forecasting tool for a Building Management System must be a flexible and robust AI solution capable of generalizing across various building types and conditions. Our study demonstrates that TS foundation models are among the most effective in tackling this challenge. Moreover, TS foundation models outperform traditional machine learning and deep learning methods in percentage-based error, particularly in SMAPE, offering a valuable metric for comparing forecasting performance across different sensor types.

However, the proper design of the pipeline is significant for deploying foundation models for building energy control and unlocking the full capability of large language models. The temporal tokenizer is one of the significant factors, as LLM was not initially proposed for processing the time series modality. LLaTA introduces an additional parallel textual branch and processes temporal inputs using a linear tokenizer, where we argue that the linear layer may be suboptimal for encoding time-series data, potentially leading to information loss. Another important observation from this study is that the single-modal foundation model (OneFitsAll) outperforms the multi-modal foundation models (LLaTA and TimeLLM). This indicates there may exist information conflicts while providing the two different modalities of data to the backbone LLM. Learned lessons from this observation is while providing LLMs with multi-dimensional information can theoretically boost the model performance to produce comprehensive predictions, the granularity and content selection of the textual description about the domain and ensuring the modalities are well-aligned are essential. Future research can concentrate on enhancing LLM's understanding of domain-specific terminology using external knowledge graphs or retrieval augmented generation techniques. This could contribute to more sustainable living environments and reduce the carbon footprint of buildings, which are significant contributors to global energy consumption and net-zero emissions.

\section{Conclusion}
\label{sec:conclusion}

This study performs a benchmark study to evaluate the TS foundation models's capabilities in building IoT forecasting. Our research investigates TS foundation models with great potential in tackling the building heterogeneity challenges, indicating the significance of temporal tokenizer selection, domain description granularity, and modality alignment. This paper also recommends future research to enhance TS foundation models' performance on building analytics with domain-specific knowledge graphs, ensuring seamless integration into real-world BMS applications while maximizing the contribution to sustainable development.



\bibliographystyle{ACM-Reference-Format}
\bibliography{0.sample-base}

\appendix

\end{document}